\documentclass[aps,prb,floatfix,showpacs,twocolumn,superscriptaddress]{revtex4}
   
\usepackage{graphicx}
\usepackage{color}
\usepackage{bm}
\usepackage{amssymb}
\usepackage{amsmath} 
\usepackage{amstext} 
\usepackage{graphics}
\usepackage{epsfig}
\usepackage{placeins}

\newcommand{\bra}[1]{\left\langle#1\right|}
\newcommand{\ket}[1]{\left|#1\right\rangle}

\def\w{\omega}

\def\e{\varepsilon}
\def\s{\sigma}
\def\S{\Sigma}
\def\r{\rho}

\def\G{\Gamma}
\def\D{\Delta}
\def\L{{\rm L}}
\def\R{{\rm R}}

\def\up{\uparrow}
\def\down{\downarrow}
\DeclareMathOperator{\Tr}{Tr}

\begin{document}

\preprint{ Version \today }

\title{Probing level renormalization by sequential transport through double quantum dots}

\author{Bernhard Wunsch}
\email{bwunsch@physnet.uni-hamburg.de}
\affiliation{I. Institut f\"ur Theoretische Physik, Universit\"at Hamburg, Jungiusstr. 9, 20355 Hamburg, Germany}
\author{Matthias Braun}
\affiliation{Institut f\"ur Theoretische Physik III, Ruhr-Universit\"at Bochum, 44780 Bochum, Germany}
\author{J\"urgen K\"onig}
\affiliation{Institut f\"ur Theoretische Physik III, Ruhr-Universit\"at Bochum, 44780 Bochum, Germany}
\author{Daniela Pfannkuche}
\affiliation{I. Institut f\"ur Theoretische Physik, Universit\"at Hamburg, Jungiusstr. 9, 20355 Hamburg, Germany}

\date{\today}

\begin{abstract}
  We study electron transport through double quantum dots in series.
  The tunnel coupling of the discrete dot levels to external leads causes a 
  shift of their energy.
  This energy renormalization affects the transport characteristics even 
  in the limit of weak dot-lead coupling, when sequential transport dominates.
  We propose an experimental setup which reveals the renormalization
  effects in either the current-voltage characteristics or in the
  stability diagram.
\end{abstract}

\date{\today}


\maketitle

\section{Introduction}\label{sec:Introduction}

Serial double quantum dots are ideal systems to 
investigate various quantum mechanical effects such as
molecular binding\cite{Rontani,chargelocalization} or
coherent dynamics\cite{Hayashi} between the constituent
dots. Furthermore, they are considered as an implementation
of a charge\cite{qbits2} or spin qubit.\cite{qbits1}
Elaborate experimental techniques were developed to
control and characterize double-dot 
structures,\cite{vanderWielreview,Exp,Waugh}
and many information about the system can be deduced from
the electric conductance through the device.\cite{Kouwenhovenreview}
Recent experiments include the measurements of quantum
mechanical level repulsion due to interdot
coupling\cite{Huttel} as well as due to external magnetic
fields,\cite{Oosterkamp} the detection of molecular states
in a double dot dimer,\cite{Blick} and the observation of
coherent time evolution of the dot states.\cite{Hayashi}

\begin{figure}
\includegraphics[width=7.5truecm]{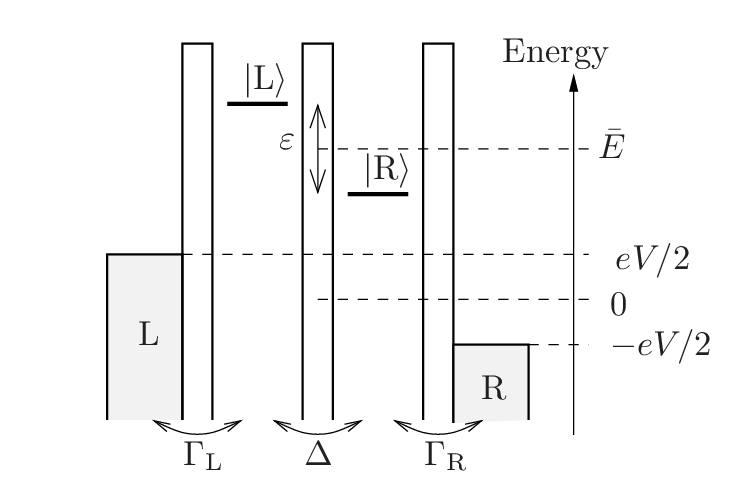}
\caption{\label{fig:model}
Schematic energy profile for a double dot coupled in series
to two reservoirs. Each reservoir is coupled to the dot of the
corresponding side by the coupling strength $\G_r$. The
inter-dot coupling is determined by $\Delta$. The energies
of the two dot states are characterized by the mean energy
$\bar{E}$ and their relative distance $\varepsilon$.}
\end{figure}

Transport through serial double dots, as depicted in Fig.~\ref{fig:model}, 
inherently visualizes the basic quantum mechanical concept of coherent
superposition of charge states.\cite{Brandes}
The states that are coupled to the left and right lead, the localized states
in the left and right dot, respectively, are no energy eigenstates of the
double dot.
This leads to oscillations of the electron in the double dot as it was shown 
in recent experiments.\cite{qbits2, Hayashi} 
To account for this internal dynamics, descriptions using classical rates 
only, are insufficient, which is why approaches including non-diagonal density 
matrix elements for the double dot have been 
developed.\cite{Nazarov1,Nazarov2,Kiesslich1,Gurvitz1,Gurvitz4}.


In this paper, we propose to use a serial double quantum
dot to probe another consequence of quantum mechanics:
the energy level renormalization of the quantum dot levels due to
tunnel coupling to a reservoir. This idea is based on two properties 
of a serial dot system. 
First, the left and right dot levels are tunnel-coupled to different
reservoirs. Since the level renormalization is a function
of the level energy, the lead chemical potential, and the tunnel coupling, 
the energy shift of the left and right dot levels is, in general, different.
Second, the conductance through the double dot is very sensitive to the 
difference of the energy levels.
It shows a resonant behavior with the width given by the tunnel 
couplings,\cite{vanderVaart} which can be much smaller than the temperature.
This sharpness of the resonance makes the conductance a valuable experimental
tool, for example to measure the shell structure of quantum
dots.\cite{Ota}


It is well known\cite{Fransson,technique1} that tunnel coupling
to reservoirs renormalizes the energy levels.
In single-dot geometries such an energy renormalization is only accessible
in transport of higher order in the tunnel coupling strength.
As we will show below, this is different for the serial double dot geometry,
for which renormalization effects are visible in the conductance already 
in the limit of weak dot-lead coupling, described by transport to first order 
(sequential tunneling) in the tunnel-coupling strength $\Gamma=
\Gamma_{\L}+\Gamma_{\R}$.

The paper is organized as follows: In
Sec.~\ref{sec:Model} we present the model Hamiltonian for the
double dot.\cite{Nazarov2, Kiesslich1, Ziegler, Gurvitz1,
Gurvitz4,Nazarov2} In Sec.~\ref{sec:kineq} we derive the
stationary density matrix and the $dc-$current for arbitrary
bias voltages. In Sec.~\ref{sec:Discussion}, we discuss how
renormalization effects appear in the $dc-$current through the
double dot.
We draw our conclusions in Sec.~\ref{sec:Conclusions}, followed by 
some technical
notes in the appendices. In App.~\ref{app:diagramaticrules}
we make explanatory notes of the diagrammatic technique,
and in App.~\ref{app:isospin} we give an illustrative
reformulation of the master equation in terms of a pseudo
spin.

\section{Model}\label{sec:Model}

We consider a double quantum dot, contacted in series, which is described 
by the Hamiltonian\cite{Nazarov2}
\begin{eqnarray}
\label{hamiltonian}
H&=&\sum_{r={\L,\R}}H_{r}+H_{\rm D}+H_{\rm T}\,.
\end{eqnarray}
The first part of the Hamiltonian describes the electric
contacts on the left $(\L)$ and right $(\R)$ side. 
These contacts are modeled by large
reservoirs of noninteracting electrons $H_{r}=
\sum_{k,\sigma} \e^{}_{rk} c^+_{rk\sigma} c^{}_{rk\sigma}$.
Here $c^{}_{r k\s},c^+_{r k\s}$ denote the annihilation and
creation operators for electrons in the reservoir
$r\in\{\L,\R\}$ with spin $\sigma$. The reservoirs are
assumed to be in equilibrium, so that they can be
characterized by the Fermi distribution
$f_{\L/\R}(\omega)$. An applied bias voltage $V$ is modeled
by different chemical potentials in the left and right contact 
$f_{\L/\R}(\omega)= f(\omega \pm eV/2)$.

The second part of the Hamiltonian, $H_{\rm D}$,
describes two dots, containing one electronic 
level each, which are coupled by the Coulomb interaction:
\begin{equation}\label{dots}
H_{\rm D} = \sum_{r={\L,\R}}E_r n_r
+ U n_{\L} n_{\R}+ U' (n_{\L \up} n_{\L\down}
+ n_{\R \up} n_{\R\down})\, .
\end{equation}
Here, $n_{i\s}=d^+_{i\s} d^{}_{i\s}$ and $n_i=\sum_\s d^+_{i\s} d^{}_{i\s}$ 
are the occupation number operators for dot $i\in\{\L,\R\}$ with spin 
$\s$, where $d_{i\s},d^+_{i\s}$ being the annihilation and creation
operators of an electron on dot $i$ with spin $\s$.
Each dot consists of a
single electronic level at the energy $E_{\L/\R}$ measured
relative to the equilibrium chemical potential of the
leads. We parameterize the levels by their average energy
$\bar{E}=(E_{\L}+E_{\R})/2$ and their difference $\e=E_{\L}-E_{\R}$, 
so that $E_{\L/\R}= \bar{E} \pm \varepsilon/2$. 
Double occupation of one individual dot is associated with the 
intradot charging energy $U^\prime$.
Simultaneous occupation of the both dots with one electron each 
costs the interdot charging energy $U$ with $U'\gg U$.
States with three or more electrons in the double dot are not considered
in the following.
The remaining eigenstates of $H_{\rm D}$, then, are:
both dots empty $\ket{0}$, one electron with spin $\sigma$
in the left $\ket{\L\sigma}$ or right dot $\ket{\R\sigma}$,
and one electron in each dot $\ket{\L\sigma \R\sigma^\prime}$.
We assume that the intra-dot charging energy always
exceeds the lead Fermi energies. Therefore the states with
two electrons in the same dot $\ket{\L\sigma \L\bar{\sigma}}$ and
$\ket{\R\sigma\R\bar{\sigma}}$ will have a vanishing occupation probability.
However, these states will appear as intermediate (virtual) states in our 
calculation, providing a natural high-energy cut-off.

The third part $H_{\rm T} = H_{\D}+H_{\G}$ of the Hamiltonian
Eq.~(\ref{hamiltonian}) describes both, tunneling between the two
dots, $H_{\D}$, as well as tunneling between dots and leads, $H_{\G}$,
\begin{eqnarray}
H_{\D}&=&- \frac{\D}{2} \sum_\s\left(\,\,d^+_{\L \s}
d^{}_{\R \s}+d^+_{\R \s} d^{}_{\L \s}\,\,\right)\\
H_{\G}&=&\sum_{k\s}
t_{\L k} c^+_{\L k \s} d^{}_{\L \s}+
t_{\R k} c^+_{\R k \s} d^{}_{\R \s}+\rm h.c.
\end{eqnarray}
Due to the serial geometry, an electron from the right
(left) reservoir can only tunnel to the right (left) dot.
The tunnel coupling of reservoir $r$ to the corresponding dot is 
characterized by the coupling strength $\G_r(\omega)=2\pi \sum_k |t_{r k}|^2
\delta(\e_{r k}-\omega)$. We consider only spin conserving
tunneling processes, and assume flat bands in the reservoirs, which
yields energy independent couplings $\G_r$.
Furthermore, we choose the interdot tunnel coupling amplitude $\D$ as a 
positive, real parameter, which can be always achieved by a proper
gauge transformation.



\section{Kinetic equation}\label{sec:kineq}
In the following section we calculate the stationary
reduced density matrix $\r_{\rm st}$ for the double dot
system and the $dc-$current through the system. 
The reduced density matrix of the double-dot is obtained from the 
density matrix of the whole system by integrating out the reservoir 
degrees of freedom.
The Liouville equation for the reduced density matrix then
has the following structure:
\begin{eqnarray}
0=i\hbar\frac{d}{dt}\bm \rho_{\rm st}
&=& [H_{\rm D},\bm \rho_{\rm st}]
   +[H_{\rm \D},\bm \rho_{\rm st}]
   +\hat{\S}\bm \rho_{\rm st}\,.
\label{master}
\end{eqnarray}
The first two parts represent the internal dynamics on the
double dot, which depends on the level separation $\e$ and
the interdot coupling $\D$. The third part of Eq.~(\ref{master})
accounts for the tunnel coupling between double dot and external reservoirs. 
The fourth order tensor $\hat{\S}$ contains imaginary and real parts,
associated with particle transfer processes and with tunnel induced 
energy renormalization of the dot levels, respectively.
The latter has been neglected in previous 
works.\cite{Gurvitz1,Gurvitz4,Nazarov2}
We calculate $\hat{\S}$ using a real-time diagrammatic 
approach\cite{technique1,technique2} as
explained in App.~\ref{app:diagramaticrules}. Also
alternative methods are available such as Bloch-Redfield
theory.\cite{Redfield, Hartmann}

In the following we concentrate on the limit of weak tunnel coupling
between double dot and leads.  Therefore, we calculate $\hat{\S}$ to
lowest order in the tunnel-coupling strength $\Gamma=\Gamma_{\rm
  L}+\Gamma_{\rm R}$, which defines the so-called sequential-tunneling
approximation.  This approximation implies that all tunneling events
are independent from each other, which is fulfilled for $k_{\rm B} T
\gg \G$. Since the correlations generated in the bath during a tunnel
process decay on the time scale $\hbar/k_{\rm B}T$,\cite{schoeller}
(this follows from the dependence of the tunneling line in
Fig.~\ref{fig:contour}-\ref{fig:diagExam} on its extension in time),
while the average time between consecutive tunneling events is given
by the inverse of the coupling strength $\hbar/\G$, higher order,
coherent tunneling events are suppressed by the condition $k_{\rm B} T
\gg \G$ and may be neglected.

The energy eigenstates of the double dot subsystem $H_{\rm D}+H_{\D}$
are the bonding and anti-bonding states with energies $E_{\rm b/\rm
  a}=\bar{E} \mp \Delta_{\rm ab}$ where $\Delta_{\rm ab}
=\sqrt{\Delta^2+\varepsilon^2}$ denotes their energy splitting.  This
identifies $\Delta_{\rm ab}$ as frequency of the charge
oscillations,\cite{Hayashi,qbits2} and $\D$ as minimum distance
between the bonding and anti-bonding eigenstates as function of the
left and right energy level.\cite{Huttel}

If the splitting exceeds the intrinsic broadening of the levels,
$\Delta_{\rm ab} \gg \G$, then the internal oscillations are fast, and
transport through the double-dot system takes place through two
separate incoherent levels.  In this case, off-diagonal matrix
elements of the stationary density matrix vanish, which can be seen
from the expansion of the Liouville equation $0=i \hbar
\frac{d}{dt}P^{\rm a}_{\rm b}= \D_{\rm ab} P^{\rm a}_{\rm b} +O(\G)$,
where $P^{\chi_1}_{\chi_2}$ denotes the
matrix element $P^{\chi_1}_{\chi_2}=\bra{\chi_1}\bm \rho_{\rm st} \ket{\chi_2}$ of the reduced density.  

The more interesting transport regime is in the opposite limit,
$\Delta_{\rm ab} \lesssim \G$, where the external coupling strongly
modifies the internal dynamics, which is captured by the off-diagonal
elements of the reduced density matrix.\cite{Gurvitz4,Nazarov2}
Combined with the validity condition for sequential tunneling, i.e.
$\G\ll k_{\rm B} T$ this implies $\Delta_{\rm ab} \ll k_{\rm B}T$,
i.e., internal oscillations are slow as compared to the time scale for
the correlations during a tunneling event.  As a consequence, the
localized states $\ket{\L\sigma}$ and $\ket{\R\sigma}$ can be used as
eigenstates of the double dot in the calculation of $\hat{\S}$, which facilitates the interpretation of the dynamics.
Technically, the condition $\Delta_{\rm ab}\lesssim \G \ll k_{\rm B}T$
means that for a consistent theory, we do not only expand $\hat{\S}$ in Eq.~(\ref{master}) to first order in $\Gamma$, but
also have to expand it to zeroth order in $\Delta_{\rm ab}$. This is
accomplished by replacing the energies $E_{\rm L/R}$ arising in the
calculation of $\hat{\S}$ by the mean level
energy $\bar{E}=(E_{\rm L}+E_{\rm R})/2$. (Therefore our formulas only
contain the Fermi functions at the average single particle level
$f_r(\bar{E})$, while energies of the order of the interdot tunneling
or the level separation are smeared by temperature).  It is worth to
point out, that by using the localized states as basis of the
$\hat{\S}$ (i.e. by calculating the transition rates
in the localized basis) one is automatically limited to the regime
$\Delta,\e \ll k_{\rm B}T$.\cite{Gurvitz1,Gurvitz4,Nazarov1,Nazarov2}

The technical details of how to calculate $\hat{\S}$ are described in
App.~\ref{app:diagramaticrules}. The master Eq.~(\ref{master}) then
must be solved under the constraint of probability normalization
$\Tr[\bm \rho_{\rm st}]=1$.  The stationary current $I$ is given by
the time derivative of the expectation value of the total number of
electrons in either the left or the right lead.  For the lowest-order
expansion used in the present context, the current can alternatively
be written in the form\cite{Nazarov2}
\begin{eqnarray}
\label{eq:current}
I&=& -e\frac{i}{\hbar}\langle[H_{\D},n_{\L}]\rangle=
-\frac{e}{\hbar}\D {\rm Im} (\sum_\s P^{\L\s}_{\R\s})\quad,
\end{eqnarray}
where ${\rm Im}$ denotes the imaginary part.
 In App.~\ref{app:isospin}
we give an analytical solution for the current as function of bias
voltage and gate voltages. In the following we discuss our results.

Instead of working with an off-diagonal density matrix one may switch 
to a pseudo spin representation of the problem. Then this double-dot
transport problem shows similarities to the
system of a quantum dot connected to ferromagnetic
leads.\cite{technique3,Hanle} This will be discussed in 
App.~\ref{app:isospin}.

\section{ Discussion}\label{sec:Discussion}
The stationary current takes the form
\begin{eqnarray}\label{current}
  I &=& \frac{e}{\hbar}
\Delta^2  \frac{A}{B^2+\e_{\rm ren}^2}\,.
\end{eqnarray}
The numerical factors $A$ and $B$ (the explicit form is given in
App.~\ref{app:isospin}) depend only on the tunnel coupling constants
$\Gamma_{\rm L}$, $\Gamma_{\rm R}$, and $\Delta$ as well as on the
Fermi distribution functions $f_{\rm L/R}(\bar{E})$ and $f_{\rm
L/R}(\bar{E}+U)$ of the left and right lead, but not on the level
energy difference $\varepsilon$. The current in Eq.~(\ref{current})
shows the well-known\cite{Nazarov2,vanderVaart} Lorentzian dependence
on the energy separation $\e_{\rm ren}$ between left and right dot
level.  However the energy separation is affected by the
renormalization of the bare localized levels:
\begin{eqnarray}
\label{e_ren}
\e_{\rm ren}&=&\e +\D E_{\rm L} - \D E_{\rm R} \, .
\end{eqnarray}
This is the central statement of the present paper. The
energy shift $\D E_r$ of the energy level in dot $r$,
caused by the external tunnel coupling, is given by
%
\begin{equation}
\label{renorm}
\D E_r =  \phi_r(\bar{E}) - 2 \phi_r(\bar{E}+U) + \phi_r(\bar{E}+U')
\end{equation}
with 
\begin{equation}
  \phi_r(\omega) =  \frac{\Gamma_{r}}{2\pi}{\rm Re}\, 
  \Psi\left(\frac{1}{2}+i\beta\frac{\omega - \mu_r}{2\pi}\right) \, .
\end{equation}
Here, ${\rm Re}$ denotes the real part, $\Psi$ is the digamma function, and $\mu_{\L/\R}=\pm eV/2$ the leads'
chemical potentials.
%
%
We want to emphasize that this energy renormalization is
not due to a capacitative but rather due to the tunnel
coupling to the reservoirs. Furthermore, it vanishes in the 
noninteracting case $U=U'=0$. The intradot charging energy
$U^\prime$ (which we usually treat as infinite to avoid
double occupation of one dot) serves as a natural cut off
for the energy renormalization in Eq.~(\ref{renorm}).
This is the reason why we allowed the intermediate states
$\chi_5$ in App.~\ref{app:diagramaticrules} to occupy these
states.

\begin{figure}[ht]
\includegraphics[width=1.0\columnwidth,angle=0]
{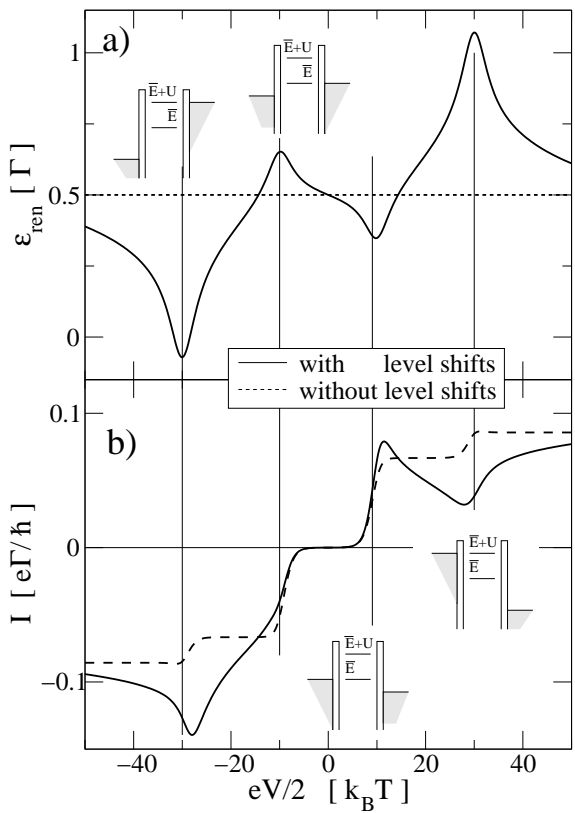}
\caption{\label{fig:I-V}
Upper Part: Renormalized level spacing $\e_{\rm ren}$ (solid
line) between the electronic levels in the left and right dot as
function of the transport voltage $V$. $\e_{\rm ren}$ is
extremal, when the chemical potential of a lead aligns with
the energy needed for either single ($\bar{E}$) or double
occupation ($\bar{E}+U$). Lower part: Current-voltage
characteristics for bare (dashed line) and renormalized level
spacing (solid line). Renormalization of energy levels
leads to an asymmetric current-voltage characteristic. The
current increases (decreases) whenever the level spacing is
reduced (increased) with respect to the bare value. Plot
parameters are: $\e=\D=\G_{\R} =\G_{\L}=\G/2$, 
$\bar{E} = 10k_{\rm B}T$,  $U = 20k_{\rm B}T$, and 
$U' = 100k_{\rm B}T$.}
\end{figure}

The energy shift of the localized levels is proportional to
the tunnel coupling strength and depends on the dot level
positions relative to the Fermi energy. 
The renormalized level separation as function of
the bias voltage is plotted in Fig.~\ref{fig:I-V}a. The
renormalized level separation $\e_{\rm ren}$ reaches a (local)
extremum each time, when the Fermi energy of a lead becomes
resonant with the energy needed for single
($\mu_r=\bar{E}$) or double occupation ($\mu_r=\bar{E}+U$).

Fig.~\ref{fig:I-V}b shows the current as function of the
transport voltage taking the level shift into account
(solid line).  By neglecting the level shifts (dashed line
in Fig.~\ref{fig:I-V}b), the current shows a typical
Coulomb staircase. The steps occur when a lead chemical
potential aligns with an electronic level in the double
dot. Since the bare energy level separation $\e$ as well as
the interdot tunneling $\D$ shall be of the order of or
smaller $\G\ge\{\D,\e\}$ and we consider $\G<k_{\rm B} T$, the
different single particle states are not resolved as
individual steps in the $I-V$ staircase. The tunnel
induced renormalization leads to additional features on the
staircase (solid line in Fig.~\ref{fig:I-V}b). Whenever the
magnitude of the renormalized level spacing grows (drops) the
current decreases (increases).  This leads to a suppression
or an enhancement of the current around the steps of
the $I-V$ characteristic, leading to regions of negative
differential conductance. The width of these feature is of
the order of the charging energy and can exceed temperature
and coupling strength significantly.

Neglecting renormalization effects and assuming symmetric coupling
 to the reservoirs ($\G_L=\G_R$), the current through the double dot is
an odd function of the transport voltage (see dashed line in
Fig.~\ref{fig:I-V}b).  This is no longer the case when renormalization
is taken into account (see solid line in Fig.~\ref{fig:I-V}b). The
reason for this asymmetry is that even though the change of asymmetry,
$\D E_{\rm L} - \D E_{\rm R}$, caused by level renormalization is
antisymmetric with respect to the bias voltage, this in not true for
the total asymmetry $\e_{ren}=\varepsilon+ \D E_{\rm L} - \D E_{\rm
  R}$ due to the non-vanishing bare splitting $\varepsilon$ (see
Fig.~\ref{fig:I-V}a). A comparable asymmetry in transport through two
coupled dots was recently observed by Ishibashi {\it et al}.\cite{Ida}
and theoretically described by Fransson {\it et al}.\cite{Fransson}
However, a negative differential conductance feature can not be
uniquely linked to such renormalization effects.  Due to interface
capacities the level positions in the left and right dot are always
affected by the transport voltage in real
experiments.\cite{vanderVaart,Pfannkuche}

\begin{figure}[ht]
\includegraphics[width=1\columnwidth,angle=0]
{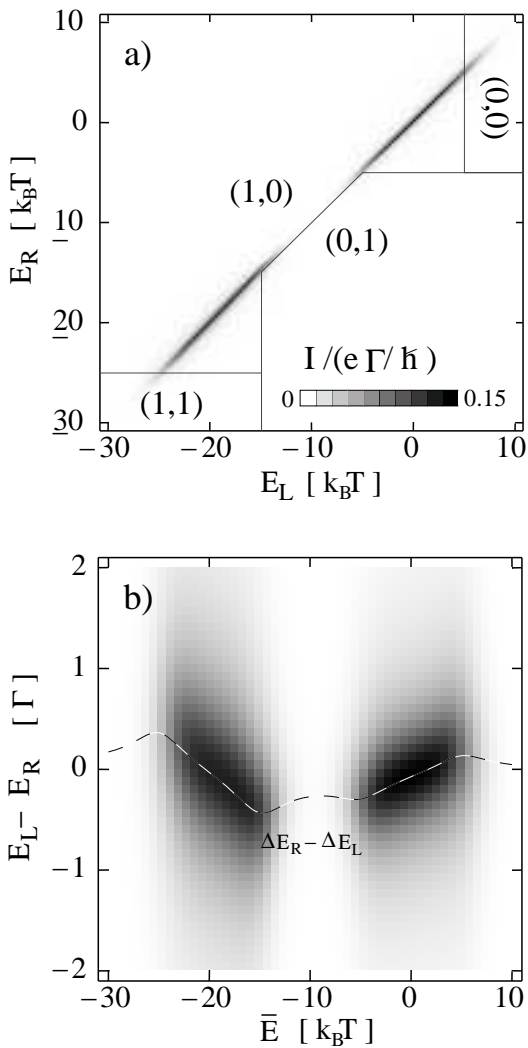}
\caption{\label{fig:e-e} Upper panel: Stability diagram
$I(E_{\L},E_{\R})$ of the current through the double dot in
the nonlinear transport regime. Well inside the areas,
separated by the black line, the occupation of the
individual dots is fixed to the written values
$(N_{\L},N_{\R})$. Elastic sequential current can cross the
structure for $E_{\L}\approx E_{\R}$ and either
$-eV/2<\bar{E}<eV/2$ or $-eV/2<\bar{E}+U<eV/2$ resulting in
two current stripes. Lower panel: Gray scale plot of the
current as function of the average single particle energy
$\bar{E}$ and the bare level separation
$\e=E_{\rm L}-E_{\rm R}$. The different renormalization of
left and right level shifts the current maxima by
$\Delta E_{\rm R}-\Delta E_{\rm L}$ (dashed black-white
line) where $\e_{\rm ren}=0$ . This leads to a tilting of
the current stripes relative to each other. Relevant plot
parameters are $k_{\rm B}T=2\Gamma$, 
$\Gamma_{\L}=\Gamma_{\R}=\Delta=\Gamma/2$,
$V=10k_{\rm B}T$, $U^\prime=100k_{\rm B}T$,
and interdot charging energy $U=20k_{\rm B}T$.}
\end{figure}

To exclude the effect of interface capacities, we propose a
different experiment: measuring the current
 $I(E_{\L},E_{\R})$ at a constant transport voltage as
function of the left and right gate voltages on the dots.
The resulting stability diagram is plotted in
Fig.~\ref{fig:e-e}a. Elastic sequential tunneling from the
left to the right dot is possible if
$E_{\rm L} \approx E_{\rm R}$. Furthermore electron
transport from the left to the right reservoir takes only
place if the dot level for single ($\bar{E}$) or double
occupation ($\bar{E}+U$) is located in the bias voltage
window. Therefore the current resonance forms two stripes
in the regions $-eV/2<\bar{E}<eV/2$ and
$-U-eV/2<\bar{E}<-U+eV/2$. Away from the
current stripes the occupation number of the left and right
dot $(N_{\L},N_{\R})$ is fixed, and no current can cross
the structure. For a detailed discussion on stability
diagrams for transport through double dots we refer to the
review of van der Wiel. {\it et al}.\cite{vanderWielreview}

In the absence of renormalization effects, the current
stripes would exactly coincide with the condition
$E_{\rm L} = E_{\rm R}$. By plotting the current as
function of the mean level position
$\bar{E}=(E_{\L}+E_{\R})/2$ and the relative energy
difference $\varepsilon=E_{\L}-E_{\R}$, one would therefore
expect a straight horizontal line. Instead, the maximum
of the current follows the renormalization shift, where the
condition $\e_{\rm ren}=0$ is fulfilled, see
Fig.~\ref{fig:e-e}b. The shift of the resonance is of
order $\Gamma$ as shown in Eq.~(\ref{renorm}) and can
be small on the scale of bias voltage or temperature.
The width of the current maxima in the stability diagram in
Fig.~\ref{fig:e-e} is not determined by temperature but
rather by the dominant coupling strength\cite{vanderVaart}
${\rm max}(\G,\D)$. Therefore the resonance width is sharp
enough to be able to measure the renormalization of energy
levels if $\G \gtrsim \D$ as used in Fig.~\ref{fig:e-e}.

In the nonlinear transport regime $\Delta E_r$ depends on
$\bar{E}$ and therefore the current stripes in
Fig.~\ref{fig:e-e} are bent and tilted against each other.
This dependence can be used as a stringent experimental prove
of the renormalization of energy levels. Due to internal
cross capacities, always appearing in real experiments, the
gate voltage of one dot is a linear function of the gate
voltage of the other dot. Therefore the stability diagram
$I(V_{\L},V_{\R})$ as plotted in Fig.~\ref{fig:e-e}a would
experience a linear shear transformation. However straight
(parallel) lines stay straight (parallel). Thus, cross
capacities can not mimic the bending due to
renormalization effects.

In real experiments in addition to the resonant current
stripes explained here, further features can arise due to
inelastic processes, cotunneling, or due to excited levels
within the bias voltage window.\cite{vanderWielreview}
These effects mainly lead to features within the triangles
below the current strips in Fig.~\ref{fig:e-e}a and are
expected not to interfere with our presented results.

Finally we compare our result obtained for the stationary
current Eq.~(\ref{currentApp}) with previous theoretical
works. For this we set the Fermi functions to
$f_{\L}(\bar{E})=1$ and $f_{\L}(\bar{E}+U)=f_{\R}(\bar{E})=f_{\R}(\bar{E}+U)=0$.
This simplifies the current to:
\begin{eqnarray*}
I=\frac{e}{\hbar} \frac{\G_{\R} \D^2}{\D^2
\left(2+\frac{\G_{\R}}{2 \G_{\L}}\right)
+4 (\e_{\rm ren})^2+\G_{\R}^2}\,,
\end{eqnarray*}
Neglecting renormalization effects (setting
$\e_{\rm ren}=\e$), this equation reproduces Eq.~(4.19) in
the paper by Gurvitz,\cite{Gurvitz4}.
Choosing the voltages such that the dot structure can also
be doubly occupied, i.e.
$f_{\L}(\bar{E}+U)=f_{\L}(\bar{E})=1$ and
$0=f_{\R}(\bar{E})=f_{\R}(\bar{E}+U)$ one obtains
Eq.~(4.18) of Ref.~\onlinecite{Gurvitz4}.

Several publications assume, that if the lead Fermi
energies are far away from the electronic states of the
dots, then the principal value integrals
(Eq.~\ref{PrinVal}), leading to the renormalization, can be
neglected.  However the energy shifts are relevant on an
energy scale given by the charging energy $U$, as shown in
Fig.~\ref{fig:I-V}a. Therefore the assumption, that one can
neglect renormalization effects and still exclude states
with more than one electron occupying the double dot is not
justified.

\section{Conclusions}\label{sec:Conclusions}
If a quantum dot is connected to a reservoirs, the tunnel coupling
causes an energy renormalization of the electronic states. We derived
the conductance of a double dot connected in series to external
reservoirs for general bias voltages and temperatures, taking into
account these energy renormalizations. We have shown, that the
conductance of such a double dot structure is affected by the energy
level shifts already in a lowest order expansion in the tunnel
coupling strength, due to its high sensitivity on the relative
detuning of energy levels. Therefore we propose to use a double-dot
system as detector for these energy renormalization effects.

We present experimental consequences of the renormalization
in the current-voltage characteristics and in the stability
diagram for the double dot in the nonlinear transport
regime. In the current-voltage characteristics we
find prominent negative differential conductances in
voltage windows of the order of the charging energy.

In the stability diagram of the double dot, we found that the current
stripes arising as function of the gate voltages for left and right
dot are tilted against each other and do not lie on a straight line,
as it is the case when energy renormalization is neglected. We showed
that the tilting of the current stripes is resolvable even in the
sequential tunneling regime (i.e. for $\Gamma>k_{\rm B}T$) as long as
the interdot tunneling, $\D$ is of the same order or smaller than the
external coupling $\G\geq \D$.

\begin{acknowledgments}
We thank S. Debald and B. Kubala, for fruitful discussions.
This work was supported by the Deutsche
Forschungsgemeinschaft via SFB~508 and under the Emmy-Noether program,
through SFB~491 and GRK~726.
\end{acknowledgments}

\begin{appendix}
\section{ Diagrammatic rules}\label{app:diagramaticrules}

With the definition $P^{\chi_1}_{\chi_2}:=\bra{\chi_1}
\r_{\rm st}\ket{\chi_2}$, the master equation
Eq.~(\ref{master}) can be written as:
\begin{eqnarray}
0=i\hbar\frac{d}{dt}P^{\chi_1}_{\chi_2}&=&
\bra{\chi_1}[H_{\rm D},\rho_{\rm st}] \ket{\chi_2}
+\bra{\chi_1}[H_{\rm \D},\rho_{\rm st}] \ket{\chi_2}\notag\\
& &+\sum_{\chi_3,\chi_4} \S^{\chi_1 \chi_3}_{\chi_2 \chi_4}
P^{\chi_3}_{\chi_4}\,.
\label{app:master}
\end{eqnarray}
In the following we show, how we calculate the tensor
$\S^{\chi_3 \chi_1}_{\chi_4 \chi_2}$, where
$\chi_i \in\{\ket{0},\rm
\ket{\L\sigma},\ket{\R\sigma},
\ket{\L\sigma\R\sigma^\prime}\}$ are the localized
eigenstates of $H_{\rm D}$, including the spin degree
of freedom.

We apply a diagrammatic technique, where each class of
tunneling processes can be represented by a diagram. Its
general derivation can be found in
Ref.~\onlinecite{technique1}. Recently this
technique was applied to Anderson-like Hamiltonians to
investigate spin-valve effects,\cite{technique3} or
signatures of the excitation spectrum in the Coulomb 
blockade.\cite{Cotunneling}

\begin{figure}[h!]
\includegraphics[width=\columnwidth]{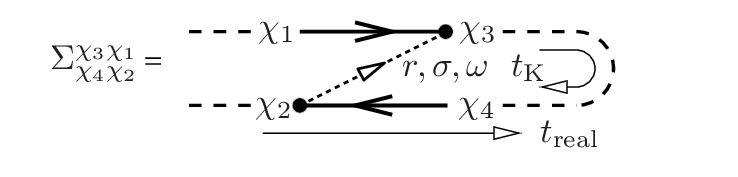}
\caption{\label{fig:contour}
Sketch of the structure of a diagram. The upper (lower)
horizontal line denotes the forward (backward) propagator
of the double dot system. The Keldysh time contour is
labeled by $t_{\rm K}$, while the real time runs from left
to right.}
\end{figure}

\begin{figure}[h!]
\includegraphics[width=\columnwidth]{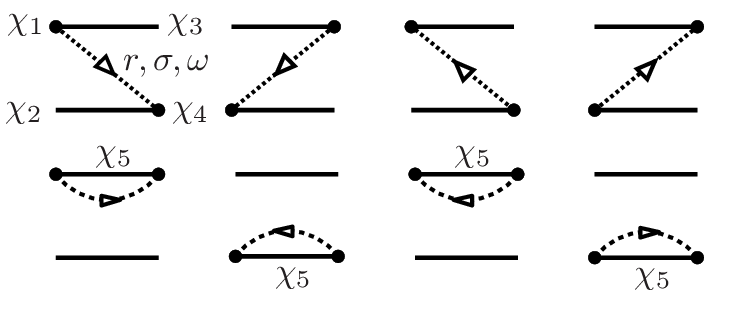}
\caption{\label{fig:alldiag}
All topologically different diagrams contributing to the
tensor $\S^{\chi_3 \chi_1}_{\chi_4 \chi_2}$ calculated in
first order in the external coupling $\G$. Labeling of the
eigenstates at the four corners and of the tunneling line
like in first diagram. $\chi_5$ labels an intermediate
charge state of the double dot.}
\end{figure}

Within this approach, the tensor
$\S^{\chi_3 \chi_1}_{\chi_4\chi_2}$ is represented as block
diagram, which is a part of the Keldysh time contour as
shown in Fig.~\ref{fig:contour}.  The upper and lower line
of the Keldysh time contour $t_{\rm K}$ represent the propagation of
the double dot system forward and backward in time. They
connect the matrix element characterized by the labels on
the left side with the matrix element characterized by the
labels on the right side. In the sequential tunneling
approximation all transitions are allowed where a single
electron first leaves and then reenters the double dot
or vice versa. The two tunnel Hamiltonians are represented by
vertices on the propagators. These vertices are connected 
by the contraction of the lead Fermi operators (indicated
by a dashed line). Each line is characterized by its energy
$\w$, the spin $\s$ of the transfered electron, as well as 
the corresponding reservoir label $r\in\{L,R\}$. A vertex
with an outgoing (incoming) tunneling line represents an
electron leaving (entering) the double dot on the specified
side $r$. All possible transitions in lowest order in the
external coupling $\G$ belong to one of the eight diagrams
depicted in Fig.~\ref{fig:alldiag}.

$\S^{\chi_3 \chi_1}_{\chi_4 \chi_2}$ is given by the sum of all
diagrams with the corresponding eigenstates at the four corners, see
Fig.~\ref{fig:alldiag}. The number of relevant diagrams is limited by
spin and particle number conservation as well as to the serial system
geometry. The rules to evaluate these diagrams in lowest order are:
\begin{enumerate}
\item{
Draw the upper and lower time contour. Add two tunnel
vertices in any topological different way. The relevant
criteria are the upper and lower contour, and the time
ordering of the vertices on the real axes, not only on the
Kelysh time contour. Assign to each free segment of the
contour a state of the double dot and the corresponding
energy. For 'bubble' diagrams like in the lower row
of Fig.~\ref{fig:alldiag}, an intermediate state $\chi_5$
participates.}
\item{
The two vertices are connected by a tunnel line. Each
tunnel line is labeled with the energy of the tunneling
electron $\omega$, its reservoir label $r$ and its spin
$\s$. Spin and reservoir label of the tunneling electron are
uniquely determined by the eigenstates involved in the
tunneling processes.}
\item{
Assign to each diagram the resolvent $1/(\D E+i0^+)$
where $\D E$ is the difference between energies belonging
to left going lines and energies belonging to right going
lines (the tunneling line as well as the propagators).}
\item{
The tunneling line connecting two vertices and labeled by
the reservoir index $r$ gives rise to the factor
\begin{eqnarray*}
\gamma_r^\pm(\omega)=\frac{1}{2\pi} \G_r f_r^\pm(\w)
\end{eqnarray*}
Here, the Fermi function
$f^+_r(\w)=f_r(\w)=1/(1+\exp[(\w-\mu_r)/k_{\rm B}T])$ corresponds to
a tunneling line that is backward directed in the Keldysh
time ordering (compare Fig.~\ref{fig:contour}), and
$f^-_r(\w)=1-f_r(\w)$ corresponds to a tunneling line
forward directed in the Keldysh time ordering.}
\item{
Each diagram gets a prefactor $(-1)^v$ where $v$ is the
number of vertices on the backward propagator. (This
leads to an $(-1)$ for the diagrams in the upper row of Fig.~\ref{fig:alldiag}.)}
\item{
Sum over possible internal eigenstates $\chi_5$ and
 integrate over the energy $\w$ of the tunneling electron.}
\end{enumerate}

\begin{figure}[h!]
\includegraphics[width=1\columnwidth]{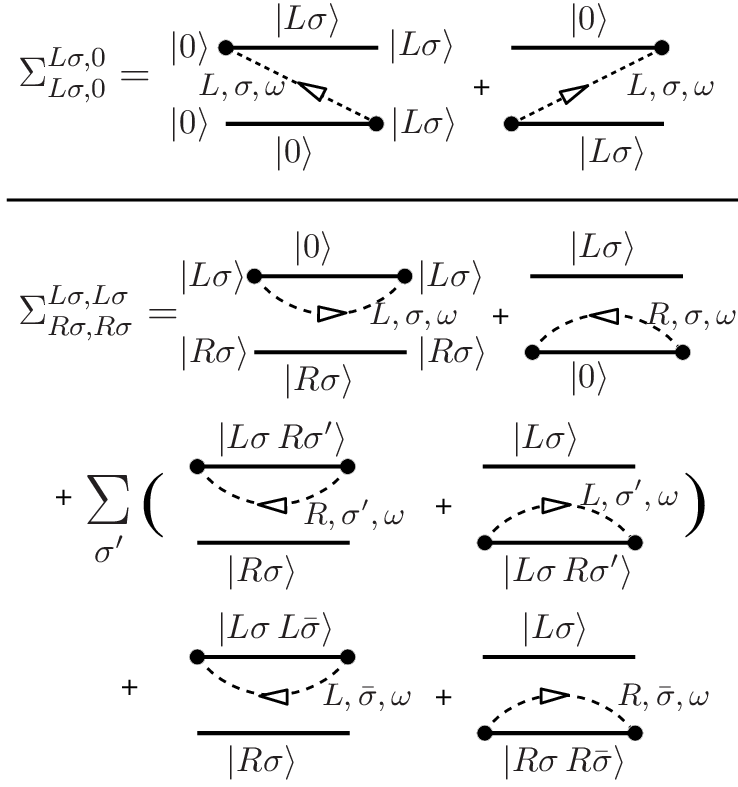}
\caption{\label{fig:diagExam}
Relevant diagrams contributing to two specific entries 
of $\hat{\S}$, in a lowest order expansion in $\G$. Every
diagram corresponding to a specific entry is labeled
by the same eigenstates at its four corners.}
\end{figure}

In the parameter regime we are interested in, the following
relations hold: $kT > \G \ge \e,\D$. Therefore the energy
difference between the single particle states is not
resolved by the Fermi functions in the reservoir, so that
we have to approximate the eigenenergies of $\{\ket{0}$,
$\ket{\L\sigma}$, $\ket{\R\sigma}$,
$\ket{\L\sigma \R\sigma^\prime}\}$ by
$\{0,E_\L \approx E_\R \approx \bar{E}, 2 \bar{E}+U\}$.
While we exclude a double occupation of a single dot for
the initial or final states by setting $f_r(\bar{E}+U')=0$
we allow the intermediate state $\chi_5$ to be in such a
state. These states have the eigenenergy $2 \bar{E}+U'$.

In Fig.~\ref{fig:diagExam}, we show as examples the diagrammatic
expansion of the tensor elements $\S_{\L\s,0}^{\L\s,0}$ and
$\S^{\L\s,\L\s}_{\R\s,\R\s}$. $\S_{\L\s,0}^{\L\s,0}$ is purely
imaginary and its magnitude has the meaning of a transition rate for a
tunneling-in process starting from the empty double dot and resulting
in a single electron with spin $\s$ sitting in the left dot. In
contrast, $\S^{\L\s,\L\s}_{\R\s,\R\s}$ also has a real part which
renormalizes the energy levels.  Calculated  in lowest order in
$\Gamma$, each element of the tensor $\hat{\Sigma}$ can
be expressed by terms of the form:
\begin{eqnarray}
\label{PrinVal2}
X_r^{(n,m)}(E)=\int\,d\omega \frac{\gamma^n_{r}(\omega)}{m(E-\omega)+i0^+}\,,\qquad
\end{eqnarray}
where $n$ and $m$ are either $(-)$ or $(+)$. In this notation, the
algebraic expression for $\S^{\L\s,\L\s}_{\R\s,\R\s}$ is:
\begin{eqnarray}\label{sigm}
\Sigma^{\L\sigma \L\sigma}_{\R\sigma \R\sigma}=&&
X_r^{(-,+)}(\bar{E})+X_r^{(-,-)}(\bar{E})\\
&&+g_\s \left(X_r^{(+,+)}(\bar{E}+U)+ X_r^{(+,-)}(\bar{E}+U)\right)\nonumber\\
&&\!\!\!\!\!\!\!\!\!\!\!\!\!\!\!
+(g_\s-1) \left(X_r^{(+,+)}(\bar{E}+U')+ X_r^{(+,-)}(\bar{E}+U')\right)\quad\nonumber
\end{eqnarray}
where, within this appendix, we allow for an arbitrary spin degeneracy
$g_{\s}$.  Since $f_r(\bar{E}+U')=0$ the imaginary part of the last
row vanishes, however this is not the case for the real part, which
causes the level renormalization.  The real part of the diagrams is
determined by the principal values of the integrals in
Eq.~(\ref{PrinVal2}) and can be expressed as a sum over digamma
functions, see Eq.~(\ref{renorm}).

Since the Hamiltonian given in Eq.~(\ref{hamiltonian}) is independent
of the orientation of the spin, each spin-realization of a charge
state is equally probable. We can therefore define $P_0=\bra{0}\r_{\rm
  st}\ket{0}$, $P^{\rm r}_{\rm r'}=\sum_{\s} \bra{\rm r\s}\r_{\rm st}
\ket{\rm r'\s}$, and $P_{\rm 2}=\sum_{\s, \s'} \bra{\L \s \R
  \s'}\r_{\rm st} \ket{\L \s \R \s'}$.  Furthermore the stationary
density matrix is diagonal in spin and particle number.  Thus the
reduced density matrix $\bm \rho$ describing the double dot is given
by the $4\times4$ matrix
\begin{eqnarray}
  \label{dotdm}
  {\bm \rho}_{\rm st}= \left(
  \begin{array}{cccc}
    P_0 &     0   &     0  & 0\\
    0   & P_{\L}^{\L} & P^{\L}_{\R}  & 0\\
    0   & P^{\R}_{\L} & P^{\R}_{\R}  & 0\\
    0   &    0    &     0  & P_2\\
  \end{array} \right) \, .
\end{eqnarray}

The diagonal elements of the density matrix are the
probabilities to find the double dot empty $(P_0)$, the
left $(P_{\L}^{\L})$ or right dot $(P_{\R}^{\R})$
singly occupied, or the two dots simultaneously occupied by
one electron $(P_2)$.  Superpositions of the two single
occupied states are possible
$P^{\L}_{\R}= \bigl(P^{\R}_{\L}\bigl)^\star$.

One can define an effective tensor for $\hat{\S}$, that only depends on
the orbital part of the matrix elements (denoted in the following
formula by $\chi_1$, $\chi_2$, $\chi_3$, $\chi_4$) and no longer on
the spin variables.  The new tensor elements are defined by:
\begin{eqnarray}
\S^{\chi_3 \chi_1}_{\chi_4 \chi_2}=\sum_{f}\S^{\chi_3^f \chi_1^i}_{\chi_4^f \chi_2^i}
\label{selfenergy}
\end{eqnarray}
Here $i$ labels any possible spin-realization for the initial states,
$\chi_1,\chi_2$, and $f$ for the final states $\chi_3,\chi_4$. (Due to
spin degeneracies the two particle states are four fold degenerate,
and the left and right states are each two-fold degenerate.)  The
tunnel tensor $\S^{\chi_3 \chi_1}_{\chi_4\chi_2}$ is independent of
the spin-realization $i$. The spin degeneracy appears only as a
prefactor, but does not change the functional form of the elements.
For example, $\Sigma_{\L,0}^{\L,0}=\sum_\s \Sigma_{\L\s,0}^{\L\s,0}$
describing the transition from $P_0$ to $P_\L$ is twice as big for
spin-degenerate electrons as for spin-less fermions. On the other hand
$\Sigma_{\L,\L}^{\L,\L}=\Sigma_{\L\up,\L\up}^{\L\up,\L\up}
+\Sigma_{\L\down,\L\up}^{\L\down,\L\up}=
\Sigma_{\L\down,\L\down}^{\L\down,\L\down}
+\Sigma_{\L\up,\L\down}^{\L\up,\L\down}$ describing the loss term of
$P_\L$ is the same for spin-degenerate or spin-less fermions since
$\Sigma_{\L\down,\L\up}^{\L\down,\L\up}=0=
\Sigma_{\L\up,\L\down}^{\L\up,\L\down}$.

This treatment  of the spin allows a general solution of
the problem including both, the case of spin polarized
electrons and the case of spin degenerate electrons. For
the interested reader, we specify the degeneracy of
fermions in the further Appendix by the variable
$g_\s$: $g_\s=2$ for electrons, $g_\s=1$
for spin-less fermions.

\section{Rewriting kinetic equation as Bloch like equation }\label{app:isospin}

Instead of working with off-diagonal density matrix
elements, we can switch to a pseudo spin
representation. As any two level system, the $2\times2$
hermitian submatrix of the singly occupied states in
Eq.~(\ref{dotdm}) can be treated as $SU(2)$ representation
of a pseudo spin Bloch vector
${\bm I}=(P^\L_\R+P^\R_\L,
i P^\L_\R-i P^\R_\L,
P_\L^\L-P^\R_\R)^T/2$.
For a complete set of variables, we further introduce
$P_1=P_\L^\L+P^\R_\R$ as the probability of a
singly-occupied double dot.
Such a pseudo spin representation is often used in the
quantum information community.\cite{qbits1,qbits2}
With this change of variables, the dynamics of the double
dot system can be mapped on the motion of a spin in an
external magnetic field. This is in close analogy to the
dynamics of a quantum dot connected to
ferromagnetic leads.\cite{technique3,Hanle}

Due to the serial geometry the external tunneling affects
only the z-direction of the pseudo spin and the left and
right contacts couple with a different sign to $I_{\rm z}$.
This is captured by the definitions
$\hat{\bf n}_{\rm L}=(0,0,1)$ and
$\hat{\bf n}_{\rm R}=(0,0,-1)$, which can be understood as
pseudo-spin magnetizations of the leads. With this
definitions the occupation probabilities obey the following
master equations:
\begin{eqnarray}
0=\frac{d}{dt} P_0&=& \sum_r \frac{\G_r}{\hbar} (-g_\s f_r(\bar{E})P_0+\frac{1}{2}f_r^-(\bar{E})P_1)+\notag\\
                & &+ \sum_r \frac{\G_r}{\hbar} f_r^-(\bar{E}) \hat{\bf n}_{\rm r}\cdot {\bf I}\\
0=\frac{d}{dt} P_2&=& \sum_r \frac{\G_r}{\hbar} (\frac{g_\s}{2} f_r(\bar{E}+U)P_1- f_r^-(\bar{E}+U) P_2)\notag\\
                & &- \sum_r \frac{\G_r}{\hbar} g_\s f_r(\bar{E}+U)\hat{\bf n}_{\rm r}\cdot {\bf I}\\
P_1&=&1-P_0-P_2
\end{eqnarray}
In equilibrium ($f_{\R}=f_{\L}$) the diagonal matrix
elements are given by the Boltzmann statistics
$P_0=1/Z,\,
P_1=2 g_\s \exp[-\bar{E}/k_{\rm B}T]/Z,\,
P_2=g_\s^2 \exp[-(\bar{E}+U)/k_{\rm B}T]/Z,\,
Z=P_0+P_1+P_2$
and the accumulation term as well as all components of the
pseudo spin vanish.

The dynamics of the single particle state is described by a
Bloch-like equation:
\begin{eqnarray}
  0=\frac{d}{dt} {\bf I}&=& \left(\frac{d {\bf I}}{dt}\right)_{\rm acc.}-\left(\frac{d {\bf I}}{dt}\right)_{\rm rel.}
                            + \frac{1}{\hbar} ({\bf B} \times {\bf I})\qquad \label{pseudospin}\\
\left(\frac{d {\bf I}}{dt}\right)_{\rm acc.}&=&
\sum_r
\hat{\bf n}_{\rm r} \frac{\G_r}{2\hbar} \left[ g_\s f_r(\bar{E}) P_0+\right.\notag\\
&+&\frac{1}{2} \left(g_\s f_r(\bar{E}+U)-f_r^-(\bar{E})\right)P_1\notag\\
&-&\left. f_r^-(\bar{E}+U) P_2\right]\notag\\
\left(\frac{d {\bf I}}{dt}\right)_{\rm rel.}&=&
\frac{1}{2}\sum_r \frac{\G_r}{\hbar}  \left(f_r^-(\bar{E})+g_\s f_r(\bar{E}+U)\right) {\bf I}\notag
\end{eqnarray}
Three different terms can be identified in the Bloch
equation. The term $\left(d {\bf I}/dt
\right)_{\rm acc.}$ describes the accumulation of pseudo
spin in z-direction due to the serial external coupling.

The relaxation-like term $(d{\bf I}/dt)_{\rm rel.}$ limits
the amount of pseudo spin. $\bm I$ relaxes isotropic by
electrons leaving or entering the singly occupied double
dot destroying all pseudo spin components.

The third term looks like a rotation of the pseudo spin
around a fictitious magnetic field $\vec{B}=(-\D,0,
\e_{\rm ren})$, where $\e_{\rm ren}$ denotes the renormalized
level separation
\begin{eqnarray}\label{PrinVal}
\e_{\rm ren}&=&(E_{\L}-
{\rm Re}\left[ X_{\L}^{(-,-)}(\bar{E})
+g_\s X_{\L}^{(+,-)}(\bar{E}+U)\right.\notag\\
&&\left.- (g_\s-1) 
X_{\L}^{(+,-)}(\bar{E}+U')\right])-({\rm L}\rightarrow{\rm R})\,,\quad
\end{eqnarray}
where ${\rm Re}$ denotes the real part. The Cauchy principal value
integrals are defined in Eq.~(\ref{PrinVal2}).

This third term describes coherent oscillations inside the
double dot which mix the accumulated spin in z-direction
with the other components. The interdot tunneling
characterized by $\D$ leads to a precession of the isospin
around the x-axes, while the energy separation between the
dot levels results in a rotation around the z-axes. It is
important to note that the renormalized level separation
between the dots changes due to the external coupling and
it is not given by the bare level separation $\e$.

\begin{figure}[h!]
\includegraphics[width=0.8\columnwidth]{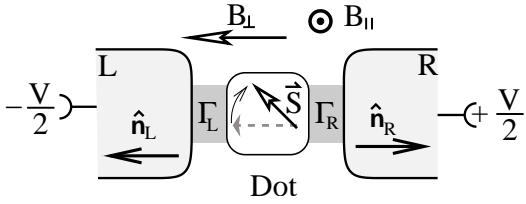}
\caption{\label{fig:spin}
Sketch of quantum dot spin valve. A single level quantum
dot is connected to two ferromagnetic reservoirs with
antiparallel magnetization. The spin precesses around an
external field with a component transverse ($B_{\perp}$)
and along ($B_{\parallel}$) the magnetization of the leads.
$B_{\parallel}$ is modified by an exchange field arising due
to the external coupling. This exchange field is manifest
in the transport properties of the spin valve.}
\end{figure}

In the following we outline the close analogy between the
transport through a serial double dot described here and
the dynamics in a spin valve described in
Ref.~\onlinecite{Hanle,technique3}.  Fig.~\ref{fig:spin}
sketches a spin valve, realized by a single level quantum
dot placed between anti-aligned ferromagnets. Relating the
pseudo spin $\vec{I}$, in the present work with the real
spin $\vec{S}$, in such a spin valve, one can perform the
following mapping.  The serial setup for the double dot
system corresponds to the anti-aligned magnetization of the
contacts in the spin valve. Furthermore the interdot
tunneling translates to a transverse magnetic field in the
single dot, while the level separation $\e$ corresponds to
the magnetic field component along the magnetization of the
contacts. Finally the renormalization of the energy levels
discussed here was introduced in the spin valve as an
exchange field leading to the Hanle effect.\cite{Hanle}

According to equation Eq.~(\ref{eq:current}) the stationary
current is just given by the $y-$component of the isospin

\begin{eqnarray}
\label{eq:current2}
I&=&\frac{i e \D}{2 \hbar} (\rho^{\L}_{\R}-\rho^{\R}_{\L})\quad=\quad\frac{e \D}{\hbar} I_{\rm y}
\end{eqnarray}
The system of master equations can be solved analytically and
the current as function of bias voltage and gate voltages has
the following form:
\begin{eqnarray}\label{currentApp}
  I\frac{\hbar}{e}&=&
\Delta^2  \frac{A}{\e_{\rm ren}^2+B^2}
\end{eqnarray}
with the factors
\begin{eqnarray*}\label{current2App}
A&=&\frac{Z_A}{N}; \quad B^2=\frac{Z_0^2}{4}+\frac{\D^2 Z_B}{N}\\
Z_0&=&(\sum_r \G_r f_{r1}^-+g_\s \G_r f_{r2})\\
N&=&g_\s\sum_r \G_r(f_{r2}^-+g_\s f_{r2})(f_{r1}^- f_{\bar{r}1}+g_\s f_{r1} f_{\bar{r}2})\\
& &+\sum_r \G_r(f_{r1}^-+g_\s f_{r1})(f_{r2}^- f_{\bar{r}1}^- + g_\s f_{r2} f_{\bar{r}2}^-) \\
Z_A&=&\frac{g_\s Z_0}{4} \left(g_\s (f_{L2}-f_{R2}) (\sum_r \G_r f_{r1})\right.\\
& &\left.+(f_{L1}-f_{R1}) (\sum_r \G_r f^-_{r2})\right)\\
Z_B&=&\frac{Z_0}{4}
\left(
\frac{\left(\sum_r \G_r^2\left(f_{r1}^- f_{r2}^-+2g_\s f_{r1} f_{r2}^-+g_\s^2 f_{r1} f_{r2}\right)\right)}{\G_L \G_R}\right.\\
& &+f_{L2}^-\left(f_{R1}^-+2g_\s f_{R1}\right)+f_{R2}^-\left(f_{L1}^-+2g_\s f_{L1}\right)\\
& &\left.+g_\s^2\left(f_{L1} f_{R2}+f_{R1} f_{L2}\right)\right)
\end{eqnarray*}
Here $\bar{r}$ denotes the opposite of $r$ and we use the abbreviations
$f_{r1}=f_r(\bar{E})$, $f_{r1}^-=1-f_{r1}$, $f_{r2}=f_r(\bar{E}+U)$,
$f_{r2}^-=1-f_{r2}$, as well as the approximation $f_r(E_L) \approx f_r(E_R)
\approx f_r(\bar{E})$.

\end{appendix}

{\small \bibliographystyle{apsrev}

}

\end{document}